\documentclass{aa}
\begin{document}

   \thesaurus{06     
              (03.11.1;  
               16.06.1;  
               19.06.1;  
               19.37.1;  
               19.53.1;  
               19.63.1)} 

             \title{AGN and starbursts at high redshift: High resolution
               EVN radio observations of the Hubble Deep Field.}


   
             \author{M.A. Garrett\inst{1}, T.W.B. Muxlow\inst{2}, S.T.
               Garrington\inst{2}, W. Alef\inst{3}, A. Alberdi\inst{4},
               H.J. van Langevelde\inst{1} T.  Venturi\inst{5}, A.G.
               Polatidis\inst{6}, K.I. Kellermann\inst{7}, W.A.
               Baan\inst{8}, A.  Kus\inst{9}, 
               P.N. Wilkinson\inst{10}, A.M.S Richards\inst{2}}

   \offprints{M.A. Garrett}

   \institute{Joint Institute for VLBI in Europe (JIVE),  
              Postbus 2, 7990~AA, Dwingeloo, The Netherlands 
              (garrett@jive.nl).
\and MERLIN/VLBI National Facility, Jodrell Bank Observatory, Lower
  Withington, Macclesfield, Cheshire SK11 9DL, UK.
\and Max-Planck-Institut f\"{u}r Radioastronomie   
  Auf dem H\"{u}gel 69,   D-53121 Bonn,  Germany.    
  \and Instituto de Astrof\'{i}sica de Andalucia, CSIC, Apdo. 3004, 
  18080 Granada, Spain.
  \and Instituto di Radioastronomia, CNR, Via P. Gobetti, 101, 
40129 - Bologna, Italy.
  \and
  Onsala Space Observatory, S-43992 Onsala, Sweden.  
  \and National Radio Astronomy Observatory, 
  520 Edgemont Road, Charlottesville, VA 22903, USA 
  \and 
  Netherlands Foundation for Research in Astronomy (ASTRON), 
  Postbus 2, 7990 AA Dwingeloo, The Netherlands.  
   \and
  Torun Centre for Astronomy,  
  Nicolas Copernicus University  
  ul. Gagarina 11, 87-100 Torun, Poland.
   \and  
  The University of Manchester, Jodrell Bank Observatory, Lower
  Withington, Macclesfield, Cheshire SK11 9DL, UK. 
 }


   \authorrunning{Garrett et al.} 
   \titlerunning{EVN Observations of AGN and Starburst systems in the HDF} 

   \maketitle

   \begin{abstract}
     
     We present deep, wide-field European VLBI Network (EVN) 1.6 GHz
     observations of the Hubble Deep Field (HDF) region with a
     resolution of 0.025 arcseconds. Above the 210~$\mu$Jy/beam
     ($5\sigma$) detection level, the EVN clearly detects two radio
     sources in a field that encompasses the HDF and part of the
     Hubble Flanking Fields (HFF).  The sources detected are:
     VLA~J123644+621133 (a z=1.013, low-luminosity FR-I radio source
     located within the HDF itself) and VLA~J123642+621331 (a dust
     enshrouded, optically faint, z=4.424 starburst system). A third
     radio source, J123646+621404, is detected at the
     $4\sigma$ level. The VLBI detections of all three sources
     suggest that most of the radio emission of these particular
     sources (including the dusty starburst) is generated by an
     embedded AGN.
 
   \end{abstract}

%

\section{Introduction}

HST observations of the Hubble Deep Field North (hereafter HDF,
\cite{W96}) have provided deep, high resolution optical images of the
distant universe. These have been vigorously followed up with
complementary radio, sub-mm, far infrared, near infrared and X-ray
observations (\cite{F00} and references therein).  The high sensitivity
radio observations of \cite{R98} (hereafter R98), \cite{M99} (hereafter
M99), \cite{R00} (hereafter R00) and \cite{G00a} have thrown new light
on the nature of the faint microJy radio source population. These
observations, spanning a range of angular resolutions
($0.15^{\prime\prime}$ to $15^{\prime\prime}$), suggest that $\sim
60$\% of faint sub-mJy and microJy radio sources have radio
luminosities, $L_{1.4 \rm GHz}<10^{23}$W/Hz, steep spectra and angular
sizes comparable or smaller than their parent galaxies. In the optical
these systems are identified with moderate redshift, star-forming
galaxies, often presenting peculiar morphologies and evidence for
interactions with other galaxies in the field. The radio emission in
these systems arises from a combination of young supernovae, supernova
remnants (SNR) and, in particular, relic SNR emission.  Approximately
20\% of the faint radio source population are identified with
relatively low-luminosity AGN, with the remaining 20\% identified with
optically faint sources with no detections down to $I=25.5^{m}$ in the
HFF and $I=28.5^{m}$ in the HDF (R98, M99, R00). The optically faint
systems are thought to be distant galaxies, obscured by dust
(\cite{R99}).  Such systems may only be detectable in the radio and
sub-mm where the effects of extinction due to dust are less
significant. Since the nature of the optical morphology and optical
spectra of these sources is unknown, it is as yet unclear whether
individual systems can be identified as pure starburst galaxies, dusty
AGN or perhaps some mixture of both phenomena. In principle, high
resolution VLBI observations can discriminate between these
possibilities.

In this paper, we present high resolution, VLBI observations
of faint radio sources in the HDF.  The aim of these pilot
observations, made with the European VLBI Network (EVN), was to probe
for the first time, the milliarcsecond (mas) radio structures of the
faint micro-Jansky radio source population. Compared with the typical VLBI
field of view, the HDF spans a relatively large area of sky and was
specifically chosen to be devoid of all strong radio sources (S$_{1.4
  \rm GHz} < 2$~mJy). The success of the observations described here
relied heavily on the recently enhanced capabilities of the EVN (in
particular the introduction of the high sensitivity MkIV data
acquisition system), together with the use of the now standard
techniques of phase-referencing (e.g. \cite{BC95}) and wide-field VLBI
imaging (\cite{G99}).

\section{EVN Observations and Data Analysis}

Observations of the HDF region were made with the European VLBI Network
on 12-14 November 1999 at 1.6 GHz. A total of 32 hours observing time
was split into two, 16 hour runs. The EVN included the 100-m Effelsberg
(DE), 76-m Lovell (UK), 32-m Medicina (IT), 32-m Noto (IT), 32-m
Cambridge (UK), and 25-m Onsala (SE) telescopes; recording data at a
total sustained bit rate of 256 Mbits/sec (including {\it both} right
(R) and left (L) hand circular polarisation). The Deep Space Network
(DSN/NASA) 70-m telescope at Robledo (ES) was also used for
approximately 11 hours of the total 32 hour run, providing data in
one polarisation (L).


The observations were made in phase-reference mode with a typical cycle
time of 2.5 minutes on the phase-calibrator and 4.5 minutes on the HDF.
The total on-source integration time on the HDF was $\sim 14$ hours.
Two phase-calibrators were used, the primary calibrator J1241+602 (a
compact $S_{1.4 \rm GHz} \sim 455$~mJy source lying $2^{\circ}$ from the
HDF) and a fainter secondary calibrator J1234+615 ($S_{\rm 1.4GHz} \sim 30$
mJy) lying only $23^{\prime}$ from the HDF). This nearby, secondary
calibrator was first detected in a targeted MERLIN survey of potential
mJy calibrator sources located within $0.5^{\circ}$ of the HDF. Having
a nearby calibrator allowed us to reduce possible phase errors
associated with ionospheric disturbances (which may have been
significant during this period of high solar activity). However, at the
time the observations were conducted, the compactness of the secondary
calibrator was not known on VLBI scales, it was thus considered prudent
to only employ it in every other tape pass, i.e.  $\sim 1/2$ the total
tape passes scheduled.

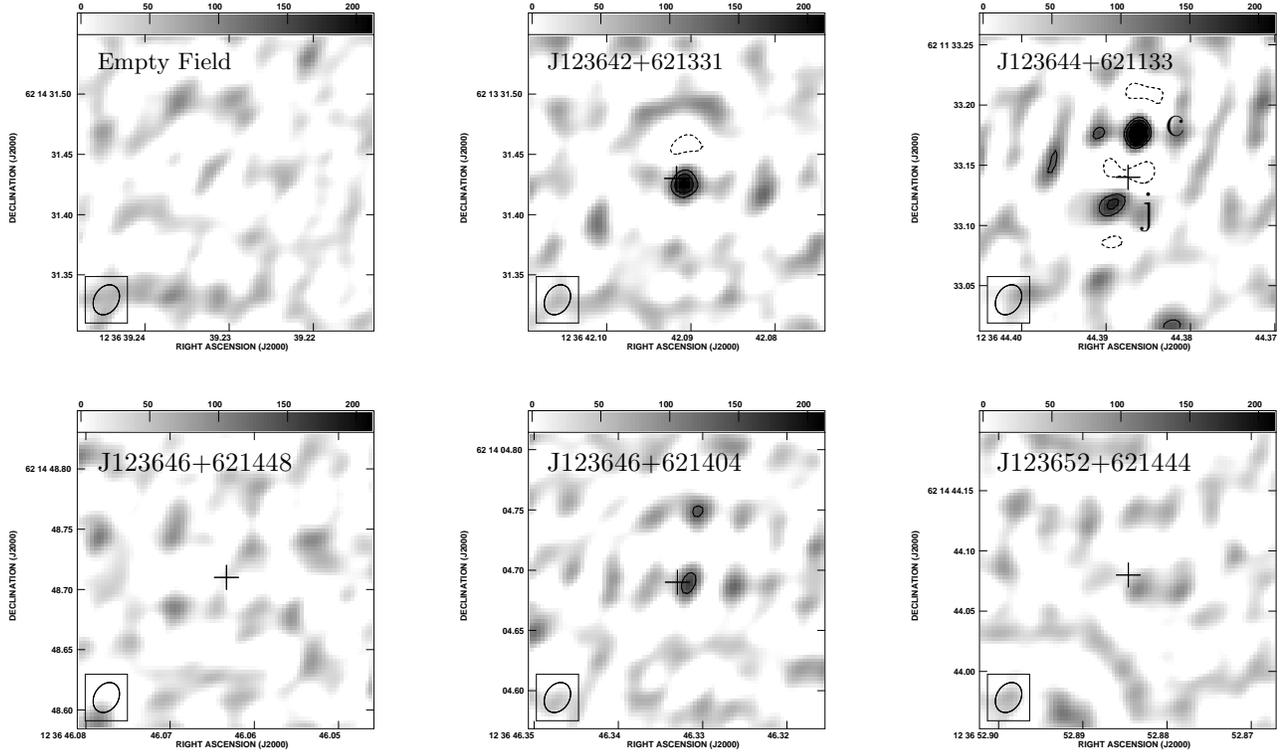
\begin{figure*}
\label{evn_hdf} 
\vspace{7cm}
\begin{picture}(150,100)

\put(0,0){\includegraphics{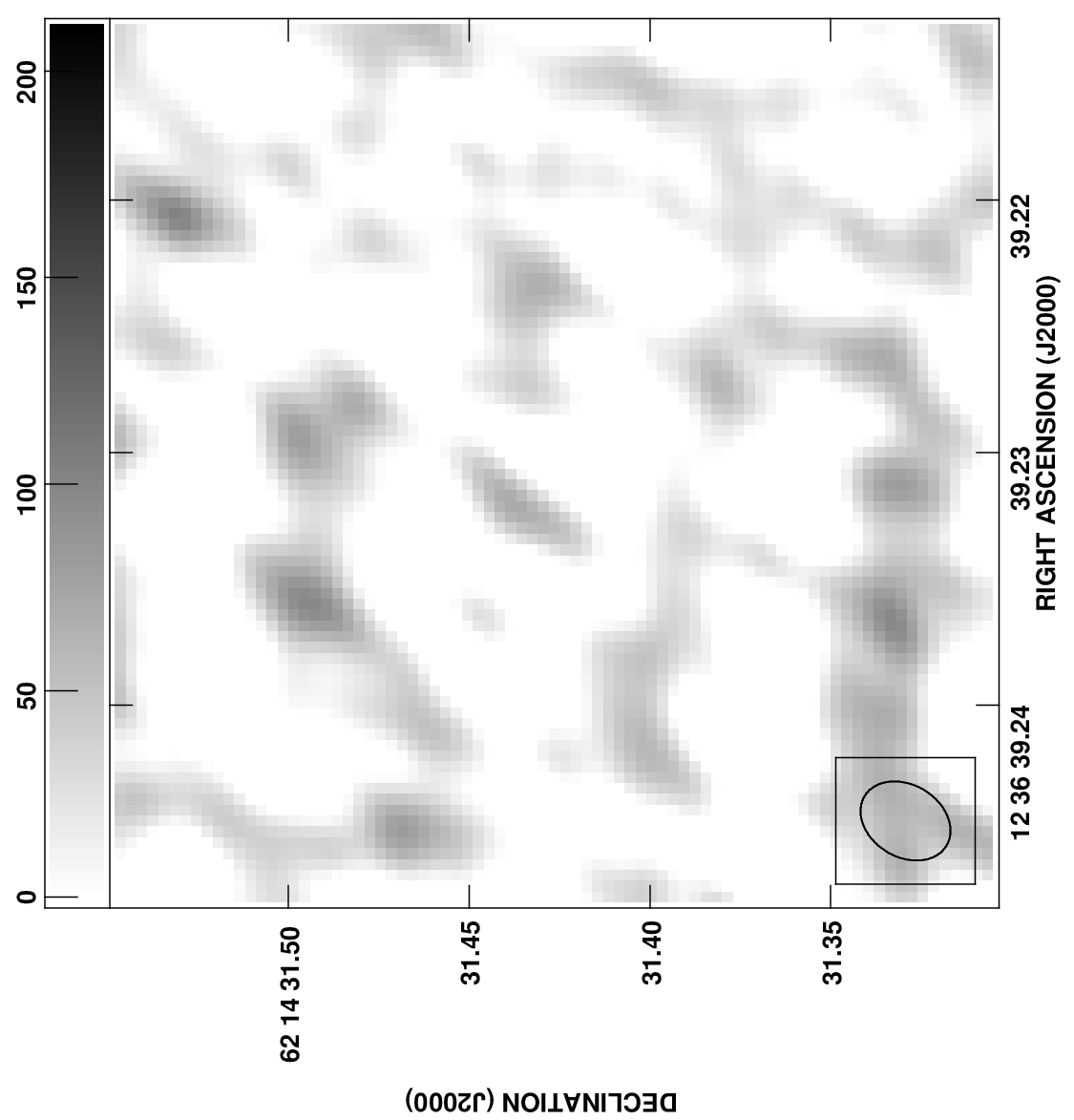}}
\put(48,265){Empty Field}

\put(0,0){\includegraphics{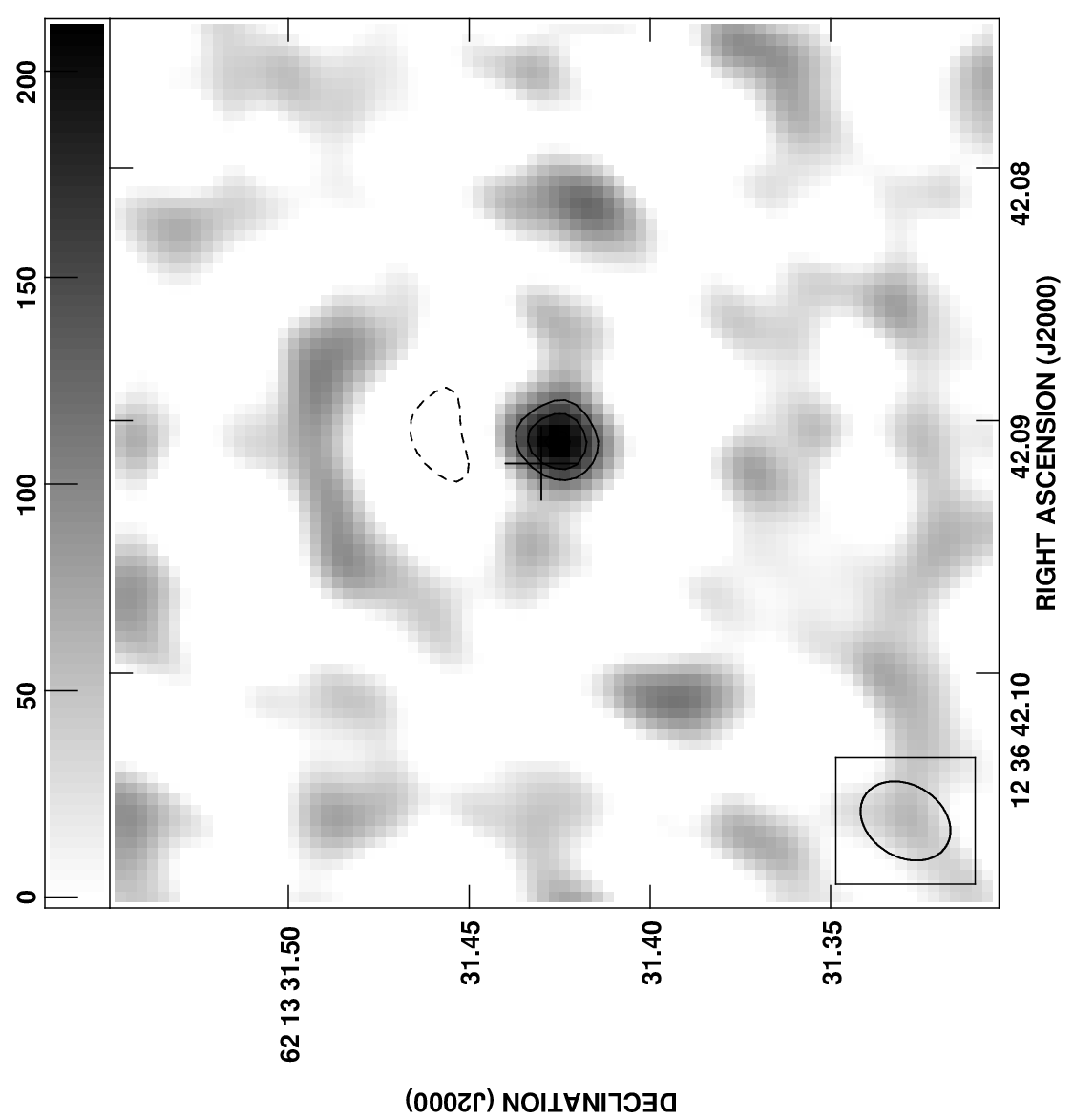}}
\put(218,265){J123642+621331}

\put(0,0){\includegraphics{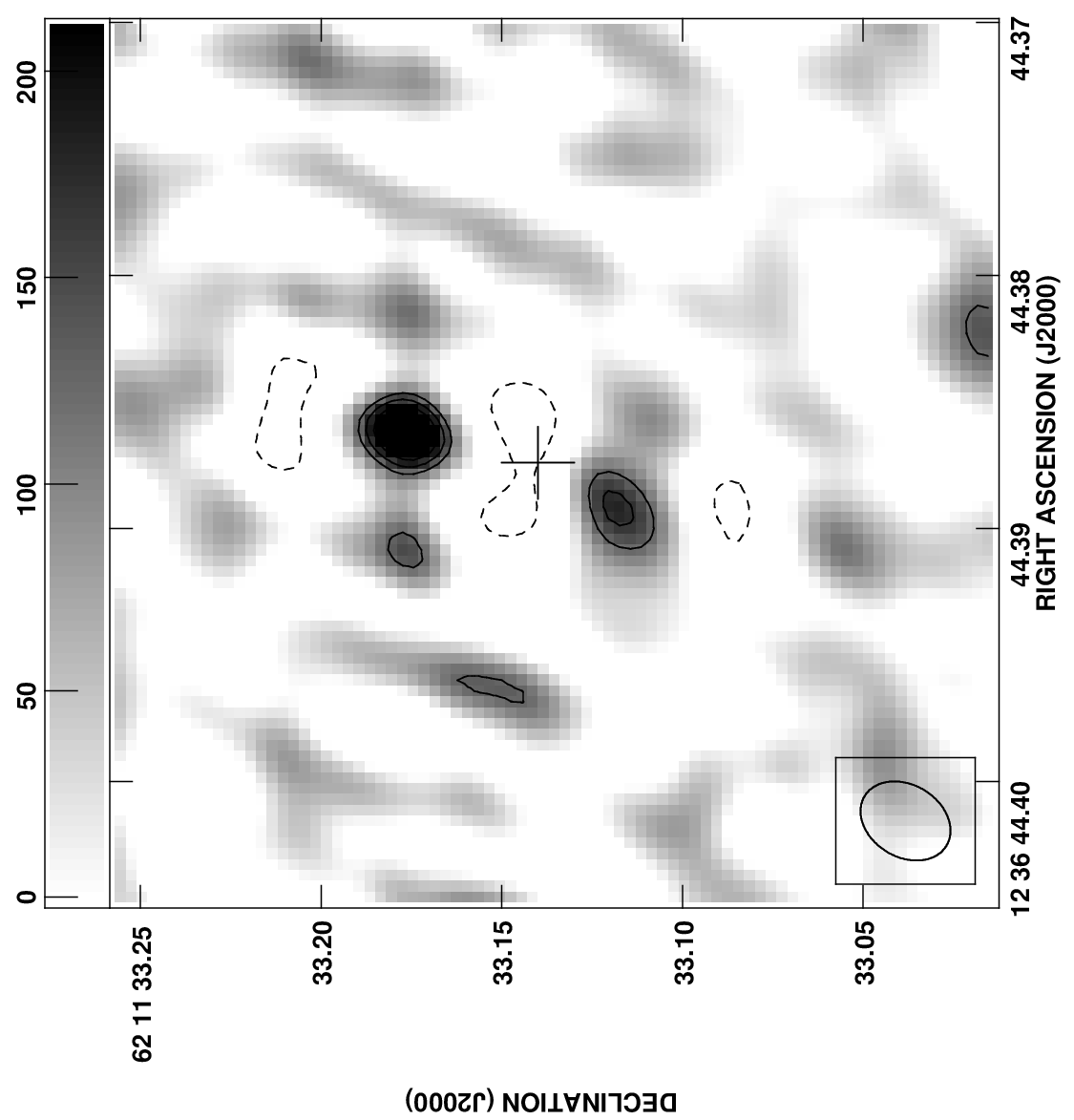}}
\put(388,265){J123644+621133}
\Large
\put(452,240){c}
\put(443,207){j}
\normalsize

\put(0,0){\includegraphics{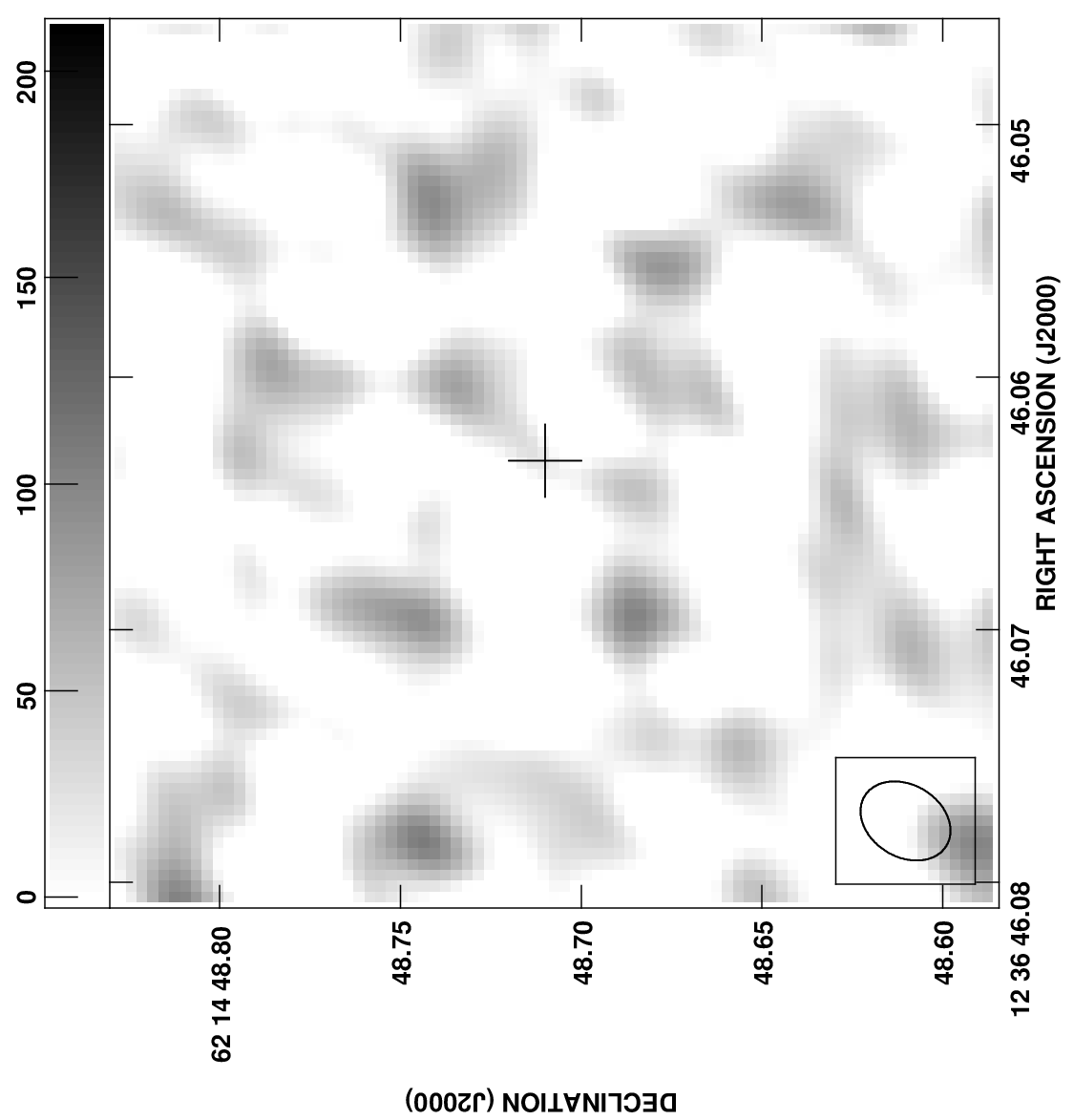}}
\put(48,113){J123646+621448}

\put(0,0){\includegraphics{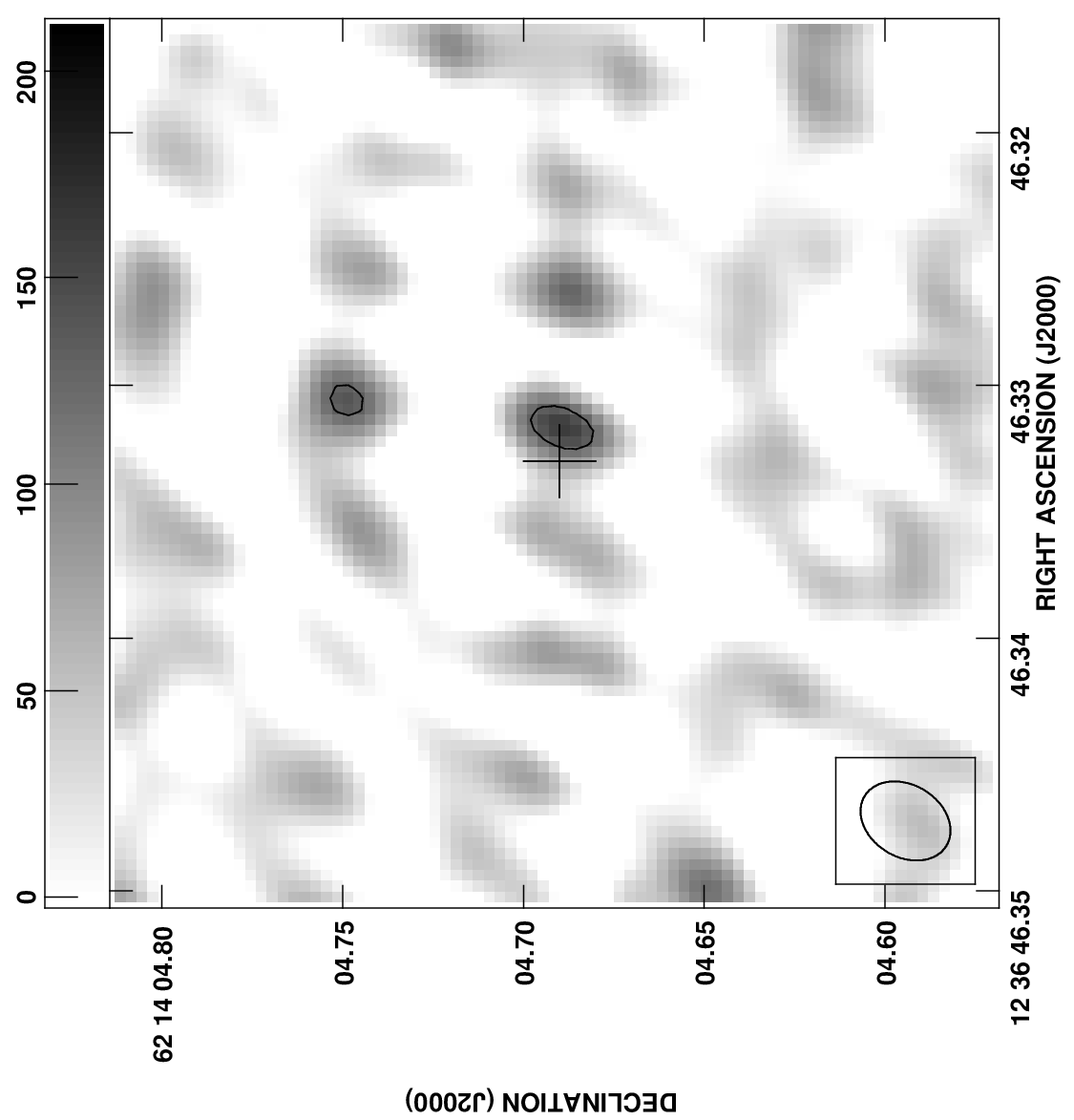}}
\put(218,113){J123646+621404}

\put(0,0){\includegraphics{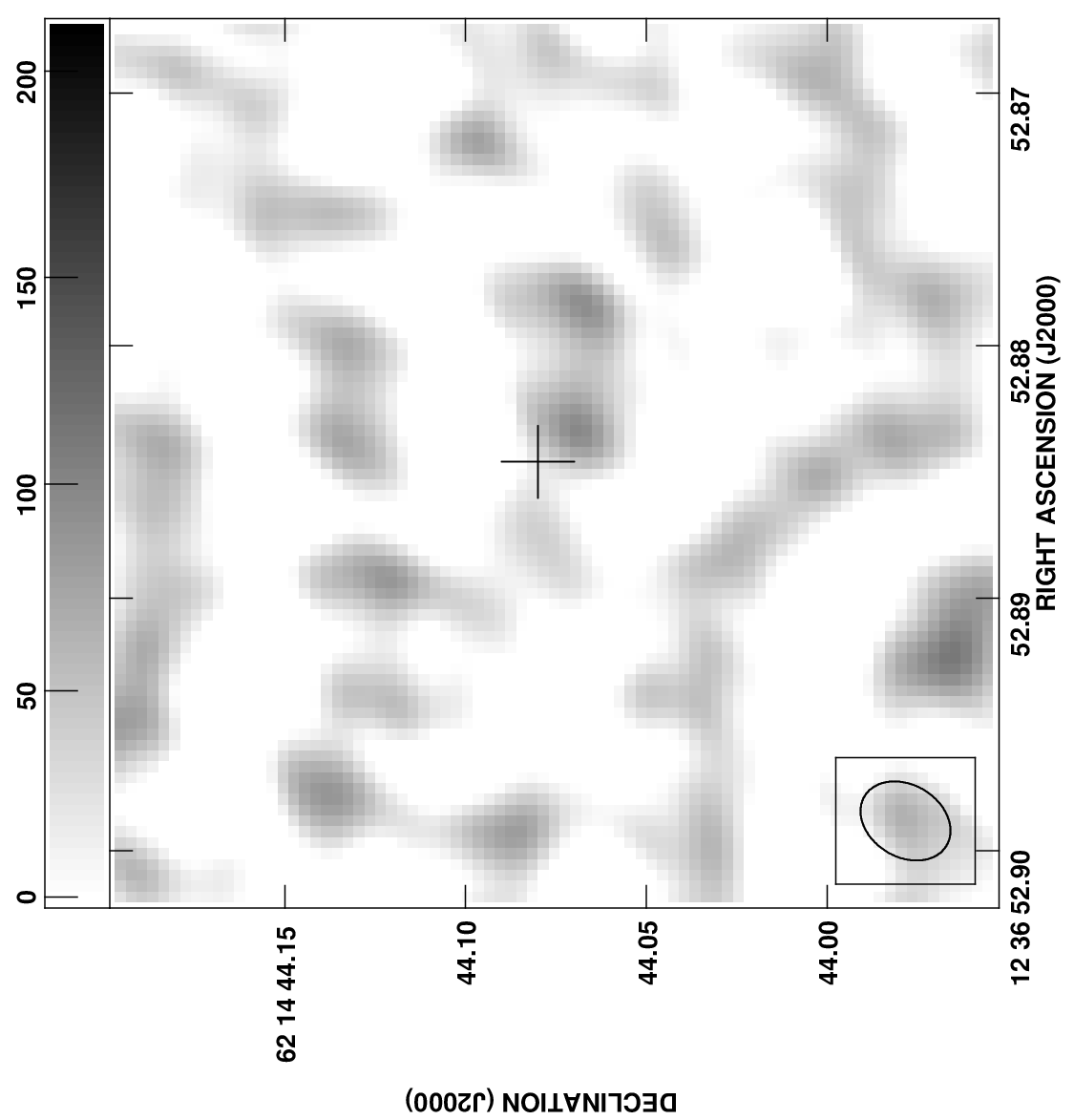}}
\put(388,113){J123652+621444}

\end{picture}
\caption{EVN 1.4 GHz dirty images of the six MERLIN targets. The
  MERLIN source positions are identified by crosses. Contours are 
drawn at -3, 3, 4, 5, 6, 7, 8, 9 times the rms noise level 
$\sim 40\mu$Jy per beam.}
\end{figure*}

The data were processed at the NRAO VLBA correlator in Socorro, NM,
USA.  The phase centre (and telescope pointing position) of the single
pass correlation was coincident with a 470~$\mu$Jy radio source,
VLA~J123642+621331 (R00), lying just outside the central HDF (in an
adjacent HFF). All subsequent data processing was performed with the
NRAO {\tt AIPS} package. The visibility amplitudes were calibrated
using the system temperatures and gain information provided by each
telescope. The residual delays, fringe rates and antenna gains were
determined from J1241+602, then interpolated and applied to the data.
The secondary calibrator was found to have compact structure on EVN
angular scales, and thus second order, phase-only corrections were also
determined for those tape passes that included the secondary
calibrator.

We selected as possible targets, all sources located within
$3.5^{\prime}$ of the EVN phase-centre with MERLIN peak flux densities,
$S_{1.4 \rm GHz}$, in excess of 60~$\mu$Jy (as measured from uniformly
weighted maps).  Five target sources satisfied these conditions, all of
which turned out to be located within $2^{\prime}$ of the phase-centre.
A sixth, ``blank'' field was also chosen, in order to act as a check on
the robustness of the imaging process. With the final gain corrections
applied, the HDF data were Fourier transformed to the image plane and
dirty images were simultaneously generated for the six target fields.

In order to image the entire HDF (an area of $\sim 5$~arcmin$^{2}$),
the HDF data were stored in the form in which the correlator generated
them: 512 contiguous 0.125 MHz spectral channels with a visibility
averaging time of 1 second. We estimate that for these data, the
effects of bandwidth and time smearing are minimal, resulting in no
more that a 10\% reduction in source intensity for fields located more
than 3$^{\prime}$ from the phase centre. A more important limitation
on the field-of-view is the fall off of the primary beam response of
the larger telescopes in the array (primarily Effelsberg, Lovell and
Robledo). The contribution made by these telescopes dominate the
naturally weighted images presented here. The most distant target
field presented in these observations is located 2 arcminutes from the
telescope pointing centre. We estimate that our response is down by
less than 15\% in fields located 2 arcminutes from the phase and
telescope pointing centre.

The naturally weighted dirty images are shown in Fig~1.  The images
were restored with a Gaussian elliptical beam ($26\times20$ mas in
PA$=-31^{\circ}$). The rms noise levels range from $42$~$\mu$Jy/beam (in
the field with the strongest detection) to $33$~$\mu$Jy/beam (in a field
with no detection). This is higher than the predicted thermal noise
($\sim 23$~$\mu$Jy/beam) which suggests we are currently limited by
errors in the calibrator phase solutions.  With the maximum side-lobe
level at 20\% of the main dirty beam response, CLEANing the 2 fields
with clear detections produced no substantial improvement in the image quality.

\section{EVN Detections in the central HDF region} 

\begin{table*}
\caption{EVN 1.4 GHz Detections in the HDF Region}
\begin{tabular}{|l|l|l|l|l|l|l|l|} 
\label{table1} 
Source & RA ($+ 12^{h}$ $36^{m}$) & DEC ($+ 62^{\circ}$) &   S$_{peak}$
& S$_{total}$ &
Size & S$_{1.4}$ & S$_{8.4}$ \\ 
       & (s)        & ($^{\prime}$ $^{\prime\prime}$) 
& $\mu$Jy & $\mu$Jy & mas & $\mu$Jy &$\mu$Jy \\ 
\hline 

VLA~J123644+621133 (c) &  
$44.38626\pm0.0001$ &  11 $33.1767\pm0.001$  
& $351\pm41$ & $ 351\pm 41$ &  
$\leq 20$ & 1290 & 477 \\ 

VLA~J123644+621133 (j) &  
$44.38933\pm0.001$ &  11 $33.1162\pm0.002$  
& $190\pm41$ & $ 214\pm77$ &  
$\leq 30$ & - & - \\ 

VLA~J123642+621331 & $42.09078\pm0.0002$ &  13 $31.4254\pm0.001$ 
& $248\pm37$ & $ 248\pm37$  & $\leq 20$ 
& 467 & 73 \\

VLA~J123646+621404 & $46.33177\pm0.0002$ &  14 $04.6893\pm0.002$ 
& $180\pm38$ & $ 180\pm38$  & $\leq 20$ 
&  179 & 168 \\

\hline
\end{tabular} 

\end{table*} 

At the $5\sigma$ detection level, the EVN observations clearly detect
VLA~J123642+621331 (at the phase centre) and VLA~J123644+621133
(located within the HDF itself). Both sources are detected
independently in both left and right-hand circular polarisation. The
sources are separated by $\sim 2^{\prime}$. There is a $4\sigma$
detection of J123646+621404, located within 20~mas of the measured
MERLIN position. The EVN source positions, flux densities and sizes of
all 3 sources is presented in Table 1.  The 1.4 and 8.4~GHz VLA flux
densities (R98, R00) of the sources are also given. Two other target
sources, J123646+621448 and J123652+621444, with peak MERLIN flux
densities of 60 and 103~$\mu$Jy were not detected ($\geq 3\sigma$
level, see Fig.~1). Note that since the primary calibrator (J1241+602)
has an ICRF position, the absolute positions of the sources presented
in table 1 are estimated to be accurate to better than 10
milliarcsecond. These are the most accurate ICRF positions associated
with the HDF region.

\subsection{VLA~J123644+621133} 

VLA~J123644+621133 is identified with a bright ($I\sim 21.05^{m}$)
z=1.013, red elliptical galaxy (R98). The radio source is clearly
identified by its large scale radio structure and luminosity as an FR-I
(R98, M99).  The EVN (see Fig.~1) detects a compact core component (c)
with an estimated size $< 20$~mas, implying a brightness temperature,
T$_{b} > 4\times10^{5}$~K. We also detect an additional $5\sigma$
component (j) located $\sim 60$~mas south of the core component.
Presumably this source (j) is a knot in the extended VLA/MERLIN jet
that also trail southwards of the core in a similar position angle.
Note that the MERLIN position for this source (see Fig.~1) is strongly
affected by the extended jet structure, lying mid-way between the
compact core (c) and the knot in the jet (j).  The VLBI structure
observed in this source, supports the view (M99) that the rare
``classical'' radio sources detected in the HDF and HFF, are the mJy
tail of the AGN radio source population.  It is worth noting that
VLA~J123644+621133 is probably an order of magnitude more distant than
any other classical FR-I radio source so far studied with VLBI.

\subsection{VLA~J123642+621331}

VLA~J123642+621331 lies just outside the HDF, in an adjacent HFF.  At
radio wavelengths it is a relatively bright, steep spectrum radio
source ($S_{1.4 \rm GHz}\sim470$~$\mu$Jy) with no optical counterpart to
$I\leq 25^{m}$ (R98). Deep HST NICMOS $ 1.6$~$\mu$m imaging clearly
detects an extremely red $23.9^{m}$ counterpart to the radio source
with a disk profile, and spectra obtained by the Keck~II telescope show
a single strong emission line, which is identified with Ly$\alpha$ at
z=4.424 (\cite{Wad99}, hereafter W99).  VLA~J123642+621331 is also
identified in the ISO supplementary list of sources
(S$_{15\mu}\sim23^{+10}_{-12}\mu$Jy, \cite{A99}) but there is no
850~$\mu$m sub-mm SCUBA detection to a limit of 2~mJy (\cite{H98}). W99
interpret the source as a dust obscured, nuclear starburst with an
embedded AGN.

The high resolution $0.15^{\prime\prime}$ 1.4~GHz VLA-MERLIN
observations (M99) resolve VLA~J123642+621331, showing that 10\% of the
flux density resides in an extended component lying to the east of an
unresolved core. It is the core which is detected by the EVN at
1.6~GHz. No other sources of radio emission ($>3\sigma$) are detected
in a $0.7^{\prime\prime}\times0.7^{\prime\prime}$ field centred on the
core.  The main question is whether the radio emission arises due to
star-formation processes or AGN activity.  The EVN limit on the
measured core size ($< 20$ mas) implies a brightness temperature,
T$_{b}>2 \times 10^{5}$K. While this in itself does not rule out a
starburst interpretation for the radio emission, the high luminosity
($\sim 10^{25}$W/Hz i.e. $\sim 100$ times more luminous than the
ultra-luminous starburst Arp 220) argues strongly for an AGN
interpretation for most of the radio emission.  Further, high
resolution, global VLBI observations are required to confirm an AGN
interpretation for this radio source.

%
%

\subsection{VLA~J123646+621404} 

VLA~J123646+621404 is a flat spectrum radio source located within the
HDF itself and is associated with a very red, face on, z=0.960 spiral
galaxy (R98). \cite{RR97} interpret the ISO 7~$\mu$m emission in terms
of a massive starburst but the presence of broad MgII emission
(\cite{P97}), radio variability (R98) and X-ray emission (\cite{H00})
suggest the spiral is host to an AGN. We see a $4\sigma$ radio source
in this field that is coincident with the MERLIN position (see
Fig.~1). We suggest this is compact radio emission from the AGN.


\section{Conclusions}

The results presented here clearly demonstrate the potential for VLBI
imaging of the micro-Jansky source population.  We have reliably
detected two radio sources within the HDF region. One of these
VLA~J123644+621133 shows all the typical properties of a low
luminosity, classical FR-I radio source. The second, perhaps more
interesting source is VLA~J123642+621331. Our detection of a
relatively compact but high luminosity radio core, further supports the
suggestion by W99 that this is a dusty, star-forming galaxy with an
embedded AGN.  We also appear to have detected radio emission from the
AGN associated with VLA~J123646+621404.

  

Global VLBI observations with 3 times the resolution and 4 times the
sensitivity of the pilot observations presented here, are now
achievable. With a linear resolution of $\sim 30$ pc at cosmological
distances, only Global VLBI can resolve nuclear star-forming regions
like those observed in Arp~220 (\cite{S98}). Targeted observations of
the optically faint radio source population (of which
VLA~J123642+621331 is just one example) may throw new light on the
nature of the associated population of dusty sub-mm sources, recently
discovered by SCUBA (\cite{H98}). In the absence of near-infrared,
optical, ultraviolet and even X-ray emission (due to dust obscuration),
future VLBI observations will be a crucial diagnostic in distinguishing
between AGN and starburst activity in these systems.

\begin{acknowledgements}
  
  We would like to thank the staff of the EVN observatories who made
  these observations possible. We also thank Lorant Sjouwerman (JIVE)
  and the staff of NRAO for supporting the correlation of these
  observations. We thank M. Eubanks (USNO) for obtaining an ICRF position for
  J1241+602. The National Radio Astronomy Observatory is a facility
  of National Science Foundation operated under cooperative agreement
  by Associated Universities, Inc.

\end{acknowledgements}


\begin{thebibliography}{}
  
\bibitem[Aussel et al. 1999]{A99} Aussel, H., Cesarsky, C.J.,
 Elbaz, D., Starck, J.L. 1999, A\&A 342, 313. 

\bibitem[Beasley \& Conway 1995]{BC95} Beasley, A.J., Conway, J.E., 1995
in J.A. Zensus, P.J. Diamond, P.J. Napier, PASPC 82, p328. 

\bibitem[Ferguson et al. 2000]{F00} Ferguson, H.C., Dickinson, M.,
  Williams, R. 2000, To appear in ARA\&A vol. 38 (astro-ph/0004319).
  
\bibitem[Garrett et al. 1999]{G99} Garrett, M.A., Porcas, R.W., Pedlar,
  et al. 1999, NewAR 43, 521 (astro-ph/9906108).

\bibitem[Garrett et al. 2000]{G00a} Garrett, M.A., de Bruyn, A.G.,
  Giroletti, M., et al. 2000, A\&A 361, L41 (astro-ph/0008509).



\bibitem[Hornschemeier et al. 2000]{H00} Hornschemeier, A. E., 
Brandt, W. N., Garmire, G. P., et al. 2000, ApJ 541, 49. 

\bibitem[Hughes et al. 1998]{H98} Hughes, D.H., Serjeant, S.,
 Dunlop, J., Rowan-Robinson, M., et al. 1998, Nature 394, 241. 

\bibitem[Muxlow et al. 1999]{M99} Muxlow, T.W.B. Wilkinson, P.N.,
  Richards, A.M.S., et al. 1999, New Astronomy Reviews, 43, 623.

\bibitem[Phillips et al. 1997]{P97} Phillips, A.C., Guzman, R., 
Gallego, J., et al. 1997, ApJ 489, 543.  

\bibitem[Richards 2000]{R00} Richards, E.A.
2000, ApJ 533, 611.

\bibitem[Richards et al. 1998]{R98} Richards, E.A., Kellermann, K.I.,
  Fomalont, E.B., et al. 1998, AJ 116, 1039
  
\bibitem[Richards et al. 1999]{R99} Richards, E.A., Fomalont, E.B.,
  Kellermann, K.I., et al. 1999, ApJ 526, L73.

\bibitem[Rowan-Robinson et al. (1997)]{RR97}Rowan-Robinson, M.,
 Mann, R.G., Oliver, S.J., Efstathiou, A., et al. 1997, MNRAS 289, 490.  

\bibitem[Smith et al. 1998]{S98}
Smith, H.E., Lonsdale, C.J., Lonsdale, C.J., Diamond, P.J.  1998, 
ApJ 493, L17. 

\bibitem[Waddington et al. 1999]{Wad99} 
Waddington, I., Windhorst, R.A., Cohen, S.H., Partridge, R.B.,
et al. 1999, ApJ 526, L77. 
  
\bibitem[Williams et al. 1996]{W96} Williams, R.E., Blacker, B,
  Dickinson, M. et al. 1996, AJ,
  112, 1335.

\end{thebibliography}
\end{document}